# A Contribution to COVID-19 Prevention through Crowd Collaboration using Conversational AI & Social Platforms

**Jawad Haqbeen**[1], **Takayuki Ito**[1], **Sofia Sahab**[1], **Rafik Hadfi**[1], **Shun Okuhara**[1], **Nasim Saba**[2], **Murtaza Hofiani**[3] and **Umar Baregzai**[4]

[1]Nagoya Institute of Technology, Japan
[2]Cheragh Medical University, Afghanistan
[3]ANPHI, Ministry of Public Health, Afghanistan
[4]Success Gate Health Organization, Afghanistan

{jawad.haqbeen@itolab.nitech.ac.jp}, {ito.takayuki, sahab.sofia, rafik.hadfi, okuhara.shun}@nitech.ac.jp
and {nasim.saba50, dr.hofiani, umarbaregzai}@gmail.com

## Abstract

COVID-19 Prevention, which combines the soft approaches and best practices for public health safety, is the only recommended solution from the health science and management society side considering the pandemic era. This process must be promoted via facilitation support to collective urban awareness programs through public dialogue and collective intelligence. Moreover, support must be provided throughout the process to perform complex public deliberation to find issues and ideas within existing approaches that can result in better approaches towards prevention. In an attempt to evaluate the validity of such claims in a conflict and COVID-19-affected country like Afghanistan, we conducted a large-scale digital social experiment using conversational AI and social platforms from an info-epidemiology and an info-veillance perspective. This served as a means to uncover an underling truth, give large-scale facilitation support, extend the soft impact of discussion to multiple sites, collect, diverge, converge and evaluate a large amount of opinions and concerns from health experts, patients and local people, deliberate on the data collected and explore collective prevention approaches of COVID-19. Finally, this paper shows that deciding a prevention measure that maximizes the probability of finding the ground truth is intrinsically difficult without utilizing the support of an AI-enabled discussion systems.

## 1 Introduction

To date, the most reliable information in Afghanistan on COVID-19 pandemic has been disseminated through the official website and traditional social outlets of the Ministry of Public Health and health organizations. However, social media platforms such as Facebook Twitter have been shown to be important in distributing information related to COVID-19-like pandemic viruses regardless of the quality and source of information in Afghanistan. However, further collection, processing and analyses of public data is required to provide the greatest social benefits. In fact, in COVID-19-affected states such as Afghanistan, where base line health data in general is non-existent, the collection of a large-scale, reliable measure of public opinion is highly important. This emerging approach for public heath sustainability cooperation aims to solve and report health issues by facilitating partnership among communities for policy-making and the good of public health.

The simple and efficient strategy to promote COVID-19 prevention is to collect and share with the public accurate and timely data and information related to the pandemic. This approach helps people understand the situation and take required measures and actions to stay healthy. Following the simplest strategy of prevention, each citizen should be given the opportunity to cooperate on every required issue. However, certain issues may prove difficult in terms of rallying full public discourse participation due to constraints such as lack of access to internet, literacy, and so on. This can be addressed to some extent by encouraging citizens with said limitations to reach out to other connected citizens with their issues. Crowd collaboration aspires to strengthen collective intelligence and utilize that intelligence for prevention in the pandemic. Ideas on any given issue gathered from members of the health society and the public can help to make better plan for the prevention. In addition, this provides the mechanism that allows less-informed citizens to receive COVID-19 prevention info-support from the crowd. Ideally, crowd collaboration via AI to identify the pros and cons, deliberate the complex data and then utilize it for COVID-19 prevention policy will be a soft contribution from info-epidemiology and info-veillance prospective.

The benefit seems to be higher when there is accurate ground truth from both public and health experts. AI aims to uncover this accurate ground truth through deliberating on the collected data and identifying the key issues and ideas. In this case, the infected people, frontline COVID-19 fighters, and the best-informed members of society will most probably provide the most-correct opinions to help and achieve prevention. Then, by employing an AI discussion system, the complex public data will be utilized for the policy-making and for the fight against the COVID-19 pandemic.

To study how effective crowd collaboration is in prevention of COIVD-19, we conducted large-scale social experiments in collective intelligence and deliberate public opinions from an info-epidemiology prospective to identify the best health experts and the public's collective concerns and intelligence on prospective prevention approaches.

This study offers a soft approach on the use of AI to identify, as per the opinions of health experts and the public, the pros and cons of COVID-19 prevention policy-making in Afghanistan. This was done by implementing an AI application for COVID-19 discussion. Given the page limit, we define the scope of this paper quite narrowly to include only the main parts of our case study.

We present the background and related use of AI for fighting COVID-19 from an info-epidemiology and prevention perspective in Section 2. Next, research aims, problem definition and methodology are discussed in Section 3. We present an overview of systems we have used during our social experiments in Section 4. In Section 5, we present the outline of the collaboration process and setting of digital social experiments. Finally, the digital social experimentation results are presented in Section 6, and the conclusion follows.

## 2 Background & Related Work

In this section, we present the background and some related recent studies in which AI has been used as a representative application in fighting the COVID-19 outbreak.

### 2.1 COVID-19 Pandemic in Afghanistan

The very first detected positive COVID-19 case in Afghanistan was confirmed in eastern Herat province on February 24, 2020. The city has long borders with Iran and the infected person also came from there to Herat province. Iran is a known COVID-19 hotspot in Asia, following China. In Afghanistan, the COVID-19 pandemic has spread all over country, and has significantly affected every aspect of life. At the time of this writing, the number of infected cases and deaths has been exponentially increasing day by day and, despite declaring a state if emergency, there is no sign that it will be under control any time soon. As of May 28, 2020, a cumulative total of 13,659 people have been diagnosed as infected with COVID-19 in Afghanistan and 246 are said to have died as a result of the virus. Kabul has had the largest share of cases at 5,093, with Herat at 2,105, Balkh at 947, Kandahar at 636 cases, and Nangarhar at 630 cases, to round out the 5 most infected states. The government directed a complete lockdown of these states and declared a state-of-emergency in the remaining 29 states. In Afghanistan, there are only 10 COVID-19 health lab which can conduct RT-PCR test; four of these are located in Kabul and the remaining six are in Herat, Balkh, Kandahar, Nangarhar, Paktiya and Kunduz provinces. After these tests help to identify the infected, they are isolated in order to prevent the outbreak. Giving this fact that early prevention is important, we conducted this study.

### 2.2 Application of AI in fighting COVID-19

One of the most effective methods to combat the COVID-19 pandemic is taking early measures to prevent outbreak. The collection and then use of reliable information, sharing, distributing and finally dissemination plays a key role for health welfare of society. The social platforms, regardless of quality and source of data, have been proven to be important in distributing information related to the COVID-19 pandemic. To confirm information reliability, the data should be collected from reliable sources or with the partnership of reliable sources, such as the Ministry of Public Health and national health organizations; further analyses can be performed if the data is collected properly. Such data can be disseminated for short and long-term objectives. In the short term, the communities will discuss, use and disseminate COVID-19 related information; whereas in the long term, the ministry will use it to create public health policy. Similar to previous pandemics, the recent emergence of COVID-19 has brought several problems under study. A large volume of research has been done using AI as a representative application in fighting the COVID-19 outbreak from info-epidemiology and infoveillance perspective.

Data accumulated via Twitter has been used to track the public behavior and examine health-seeking and public reactions towards outbreak [Ganasegeran and Abdulrahman, 2020]. Also, data has also been collected from three social platforms in China to assess public concerns and risk perception, as well as to track public behavior in response to the COVID-19 outbreak [Hou et al, 2020]. It is also possible to oral communication from social media and analyze speech using AI computer applications to contribute to research on the COVID-19 pandemic [Schuller et al, 2020].

## 3 Aims and Methodology

Our action research is based on the soft system methodology presented by experimental transfer of AI technology for Info-epidemiology and info-veillance through partnership between developed and developing countries. Within this, we proposed a method to use joint functions of social platforms to redirect discussion into online AI discussion system to conduct large-scale smart COVID-19 experimentation. This is the very first time in developing county, that a national web-dialogues with Industry-Academia-Government partnership hosted by AI-based web discusison on management of COVID-19 prevention for prevention policy making.

Our paper particularly focuses on the digital social experiments of a proposed method usage for soft approach and management of COVID-19 prevention in Afghanistan. This comes from the opportunity to receive an experimental transfer of the proposed method through a partnership between Nagoya Institute of Technology, Success Gate Health Organization, Cheragh Medical University, National Public Health Institute and Fighting against COVID-19 Department of Ministry of Public Health, Afghanistan.

### 3.1 Motivation

We propose that cross-collaboration, cooperation and partnership between industry, academia, and government (I-A-G) can significantly reduce risks and prevent the outbreak. Therefore, to collect, distribute and disseminate COVID-19 related information to communities, cross platforms collaboration are greatly needed. Motivated by recent I-A-G collaboration we applied AI for large-scale COVID-related complex discussions in collective intelligence using joint functionalities of web discussion support systems and social media platforms.

### 3.2 Problem Definition

In health sciences and management, to ensure the safety of individuals or public health, two major approaches exist to deal with diseases, namely prevention and treatment. The most efficient, recommended and desirable side of the approach coin is prevention as the treatment side is expensive and will result in adverse side effects. Treatment, thus, should in general only be activated if the prevention has not been properly applied. In the case of the COVID-19 pandemic, however, this has somewhat been reversed as prevention is being called for due to the ineffective and limited treatment available. Therefore, this soft solution of prevention and control of this pandemic has been addressed through calls for social distancing and conducting public awareness sessions aimed at improving levels of prevention understanding. Beyond this, some smart tools are being employed to sort and examine the complex and vast amount of COVID-19 data for health outreach policy and decision-making. As a powerful tool to deal with a vast amount of data, AI is considered the best choice to better understand social network dynamics related to public health and to improve the COVID-19 situation. Therefore, there is great importance in the role of crowd and cross social platforms collaboration with an AI-based online discussion system to collect, deliberate, analyze, visualize and share the information. These are needed to harness the collective intelligence for social good.

## 4 Conversational AI System for COVID-19 social experiment

We used an online AI discussion system and social media platforms for COVID-related large-scale opinion collection to strengthen COVID-19 prevention strategy and policy-making.

### 4.1 Online AI Discussion System "D-Agree"

D-Agree is an online discussion system that hosts large-scale discussions. Participants can submit their opinions as text [Sengoku et al, 2016].

D-Agree is the upgraded version of CollAgree developed by our team. The main difference between CollAgree and D-Agree is the server management side, where in CollAgree it is based on a local computer and in D-Agree it is on the cloud [Ito et al, 2014].

The system adopted the IBIS model as a discussion framework [Kunz and Rittel, 1970]. It employs and extracts users' posted items and for each user posted data, a set of features is learned automatically via a discussion structure module. After that, classification will be applied based on the IBIS structure nature. Then, output will be visualized to predicate posted item traits and behavior as issues, ideas, pros and cons. The main reason to adopt IBIS was to lead the extension of argumentative discussion structure through which people clarify issues, ideas, merits and demerits of discussion by themselves for SDGs promotion [Lawrence and Reed, 2017]. Figure 1 presents the discussion structure of our system.

In our previous research, the system has been utilized to collect large-scale public opinions for Kabul city urban policy making [Haqbeen et al, 2020] and [Haqbeen et al, 2020], where in this research by customizing some features of agent we conduct COVID-19 experimentation for public health policy-making and identifying the issues within the prevention strategy in Afghanistan.

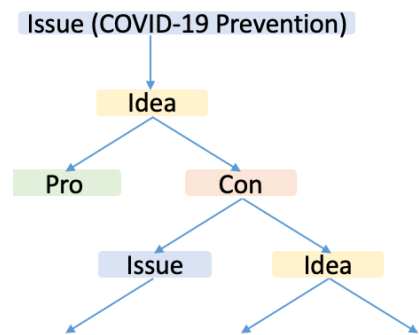

Fig.1. System's discussion structure, adopted from IBIS

### 4.2 System Architecture and User Interface

Figure 3. presents the system's general architecture. We have used Amazon Web Services to maintain scalability of large-scale discussion operation. Fig. 2 presents the user interface of our system in which we have posting, thread, theme, information media, and point areas. The post form is used to input an opinion for collecting people's opinions. Once the discussion topic has been posted, a user can reply to the topic's post form or submit opinions under the topic. In this way, the discussion leads to structure and creates a thread area. Also, a user can select the satisfaction level from 1-10 and input her/his opinion in a topic's post form or submitted opinions. The range from 1-5 represent an opposing view and 6-10 shows levels of agreement. Theme is the top-level issue and is usually defined by an administrator while creating the discussion. In our experiments, the administrators were the Ministry of Public Health and Success Gate Health organization. The virtual incentive has been designed to motivate a discussant to actively engage in discussion by giving points to his/her post, encouraging replies and likes, and their post evaluated or replied to by others [akahashi et al, 2016].

### 4.3 Automated Facilitation Agent

We developed a agenral sense automated facilitation agent that observes user-generated data, extracts their semantic discussion structures, generates facilitation messages, preserves and maintains the health of discussion by filtering out inappropriate data, classifies the posted items and visualizes the categorization of debaters' posted items. Two main mechanisms have been considered while designing the agent, discussion extraction/visualization mechanism and observing and posting mechanism.

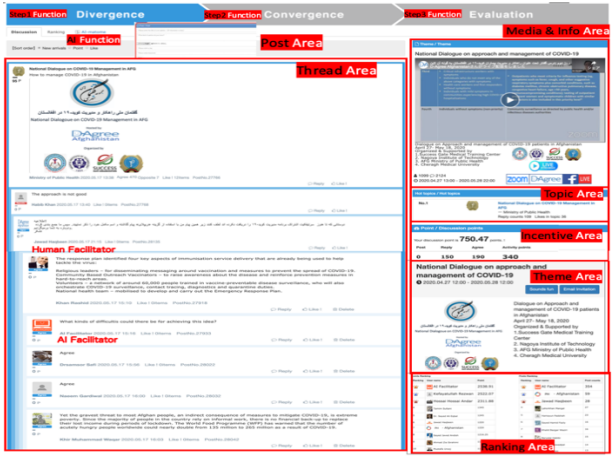

Figure 2: System user interface during social experiment

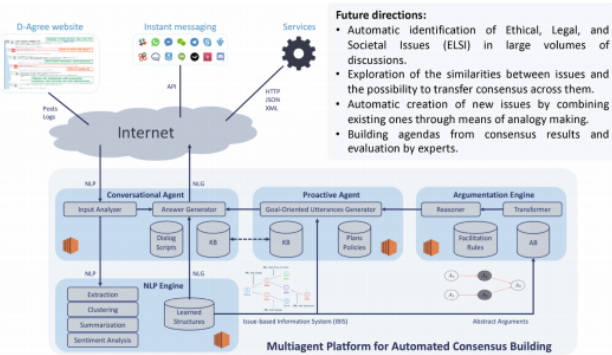

Figure 3: System general architecture

Argumentation mining technologies with BiLSTM have been utilized for the discussion extraction functionality of our agent [Stab and Gurevych, 2017]. We adopt IBIS to classify the meaningful sentences within four components, namely issues, ideas, pros, and cons. Next, the system identifies the relationship between four components and unifies these components into one discussion structure, and finally visualizes it. After that, considering the extracted discussion structure, the agent activates its observing and posting mechanism to facilitate the discussion by posting facilitation messages. The posting mechanism of our system applies 200 facilitation rules. By applying these rules and the obtained structure, the agent generates facilitation messages.

### 4.4 Facilitation Agent for COVID-19

We realized that it would be good to develop a domain specific question/answer agent with general facilitation agent for COVID-19 discussion. We proposed a COVID-19-based facilitation template, given-issue and learning dataset. We extract COVID-19 fundamental discussion data to make a COVID-19 domain dataset to train facilitation agents. Though this, we create a COVID-19 basic database for discussants to investigate and refer to past discussion by searching keywords.

### 4.5 Redirecting Social Platforms into Discussion Rooms

We focus on a novel method to guide the flow of social media and redirect the discussion into an online AI discussion system rooms using social login of our system. By doing so, we wanted to redirect and lead the flow of the aforementioned social platforms to the web discussion system and employ posted items of debaters to find truly COIVD-19 collective intelligence from public concerns, ideas and issues in response to the COVID-19 outbreak and contribute to prevention of COVID-19. By this, we want to respond to and mitigate the threat posed by COVID-19 in Afghanistan by collecting and deliberating large-scale COVID-19-related information and converge the sites into a discussion-oriented site. Table 1. shows the outline of participation and engagement level within four systems. In our experiment, D-Agree stands at the top by discussant level while Facebook maintains the first position from an attendee-level perspective. The registrants refers to those people whom actively engaged with the discussion, and attendees in ZOOM, Facebook and YouTube refers to view-only participants with discussion.

| System | Registrants | Attendees |
| --- | --- | --- |
| D-Agree | 1101 | 1.1K |
| ZOOM | 643 | 1.3K |
| Facebook | 985 | 41.1K |
| YouTube | 105 | 11.1K |
| Total | 2834 | 54.6K |

Table 1: Number of logged in users and viewers

## 5 Collaboration and Experimentation Setup

In this section, we present the collaboration setting, data collection and deliberation during our social experiments. We set up three types of audience for our social experiments namely, health experts and frontline COVID-19 fighters, local citizens and COVID-19 patients.

### 5.1 Collaboration Setting

In April 2020, Nagoya Institute of Technology launched I-A-G collaboration with Afghanistan's Ministry of Public Health, Success Gate Health organization and Cheragh Medical

University to examine the applicability of the crowd discussion system for public health discussion to drive public insights from COVID-19 dialogues in Afghanistan. This research project aims to create a sustainable crowd discussion platform by using an AI-enabled web discussion support system towards a creative smart COVID-19 dialogue to find truly social collective intelligence for public health policy and decision-making in Afghanistan.

### 5.2 Data Collection and Deliberation

We conducted a large-scale smart social experiment by integrating a live streaming function of webinars into Facebook and YouTube and then embedding it within the media platform of our AI-enabled web discussion support system. By doing so we had two main objectives. First, we lead the flow of the aforementioned social platforms into the web discussion system to collect public opinions at large. Next, we employ collected and posted items of debaters to acquire COIVD-19 collective intelligence from public concerns, ideas and issues in response to the outbreak and contribute to prevention of COVID-19.

The second objective was to extend the impact of reliable discussion out of discussion site to other platforms by converging the sites in the discussion site to use large-scale opinions in collective intelligence. Specifically, we used Kabul city's official Facebook page and YouTube to promote D-Agree and lead the social flow of these platforms towards D-Agree. Furthermore, we boosted the experiment post by Facebook ads to promote the social coverage. By this, the data flow is directed from social platforms to the discussion system and users can use their accounts on the outside systems to login and join discussion boards. A web server component manages the board and stores all data in the database for the user to refer, if they would like to disseminate the information during discussion or in the future. Next, the discussion structure extraction module of the agent extracts the IBIS structure from the discussion board while maintaining to preserve the health of discussion by observing the board and leading and integrating the discussion through facilitation. We utilized AWS Cloud Watch and Lambda functions for observation and posting module of our agent.

## 6 Social Experimental Results

We conducted a real COVID-19 social experiment with the partnership of the Ministry of Public Health from April 27 to May 29, 2020, where 588 COVID-19 frontline health experts and fighters, 492 citizens, and 21 COVID-19 patients discussed three themes about COVID-19 management and their experiences. Themes 1, 2 and 3 were created namely by Success Gate Health organization, the National Institute of Public Health and the Ministry of Public Health's COVID-19 Fighting Department, respectively. We got 3,245-page views, visits from 1,101 registered participants and a total of 2,046 opinions were submitted to the discussion board from three types of discussants, namely health experts/front line fighters, local citizens and patients. Specifically, 672 posted items were from health experts, 900 opinions were from local citizens and 48 opinions were posted by COVID-19 patients from the Afghan-Japan hospital. The Ministry of Public Health's human facilitator and our system AI agent collaboratively posted 426 facilitate messages, where 59 were by human and 367 were done by the AI agent. We established 7 time-based discussion phases, where six of them were for four days and the last one was for 5 days. We conducted an online questionnaire to check the satisfaction level of users, and we found the collaborative facilitation by human and agent successfully facilitated the discussion.

In this experiment, the number of identified opinions classified as issues were 912 items, ideas (102), merit or acknowledgement to the work done to manage COVID-19 in AFG (95), the problem within management or strategy policy of COVID-19 in AFG (184) and others (367) were extracted by the system's AI function. We found that out of 1620 net-submitted opinions, jointly submitted by experts, citizens, and patients, a total 912 of them were identified as issues, meaning 50+ submitted opinions were about issues and it is because Afghanistan is not only affected with COVID-19 but the country is conflict-affected as well. Picture (a) in Fig. 4, represents the outline of submitted opinions and picture (b) outlines the submitted opinions per each phase.

| Discussion (Weighted base) | Posting (issue) | Posting (idea) | Posting (merit) | Posting (demerit) | Posting (N/A) | Posting (Agent-F) | Posting (H-FA) | **Posting (All)** | Participants (D-Agree) |
|---|---|---|---|---|---|---|---|---|---|
| No.1 Experts | 387 | 70 | 67 | 67 | 81 | 207 | 31 | 910 | 588 |
| No.2 Local | 502 | 29 | 25 | 106 | 238 | 153 | 21 | 1074 | 492 |
| No.3 Patients | 23 | 3 | 3 | 11 | 8 | 7 | 7 | 62 | 21 |
| Total | 912 | 102 | 95 | 184 | 327 | 367 | 59 | 2046 | 1101 |

(a) Outline of extracted opinions per theme

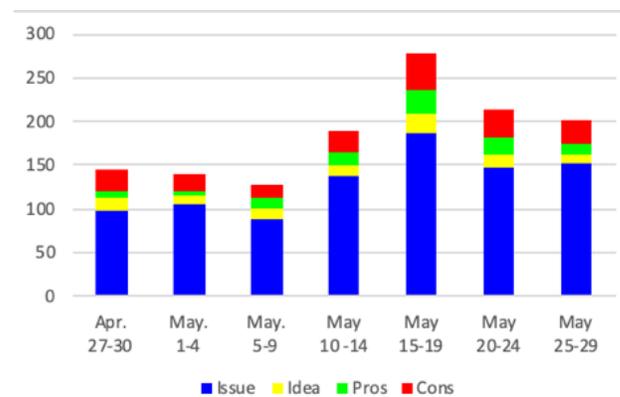

(b) Outline of extracted opinions per phase

Figure 4: Preliminary outline of discussion extraction per theme (a) and phase (b)

## 7 Conclusion

In this paper, we have reported the case study of large-scale COVID-19 social experimentation in Afghanistan using an online AI system and social platform to gather COVID-19 insights and identify the issues and ideas for public health policy making and strengthening of prevention in Afghanistan.

Firstly, we provided an introduction, the background and related work of AI for finding effective prevention approaches that can effectively help to combat the COVID-19 pandemic from info-epidemiology` of experts and local citizens concerns perspective. Then, we have presented the aims, methodology, problem definition, and motivation behind our work. Furthermore, we have discussed the collaboration between developed and developing world partnership and social experiments processes setting. Finally, we have reported and discussed the outline of preliminary experimental results and outline the impact of using AI to deliberate public discussion for COVID-19 prevention policy-making in Afghanistan. We will compare the impact of using AI in public deliberation versus not using AI in our future work.

## Acknowledgments

This work was supported by JST CREST fund Japan (Grant Number: JPMJCR15EI).